\documentclass[twocolumn,aps,showpacs,amssymb,floatfix]{revtex4}
\input epsfig.sty
\newcommand{\be}{\begin{equation}}
\newcommand{\ee}{\end{equation}}
\newcommand{\beq}{\begin{eqnarray}}
\newcommand{\eeq}{\end{eqnarray}}
\usepackage{bm}
\usepackage{graphicx}%

\begin{document}
\setcounter{figure}{\arabic{figure}}

\title{ The static tetraquark and pentaquark potentials }
\author{C.~Alexandrou and G.~Koutsou}
\affiliation{Department of Physics, University of Cyprus, CY-1678 Nicosia,
 Cyprus}

\date{\today}%

\begin{abstract}
We evaluate the static $qq\bar{q}\bar{q}$ and $qqqq\bar{q}$
 potentials in the quenched theory
at $\beta=5.8$ and $\beta=6.0$ on a lattice of size $16^3\times 32$. We
 compare the static potentials
to the sum of two meson potentials for the tetraquark system and to
the sum of the baryonic and mesonic potentials for the pentaquark state, 
as well as, with the confining potential obtained in the strong 
coupling expansion. 
\end{abstract}

\pacs{11.15.Ha, 12.38.Gc, 12.38.Aw, 12.38.-t, 14.70.Dj}

\maketitle


\section{Introduction}

A large amount of effort has been devoted recently to experimental 
searches~\cite{experiment}  
for the identification  of the $\Theta^+$, 
an exotic baryon state with an unusually  narrow width,
which was predicted theoretically in the chiral soliton 
model~\cite{DPP}. Several studies  in lattice QCD have looked
for such a state in order to
determine its mass and parity but no consensus
has been reached yet with some groups
reporting  a bound state with mass close 
to the experimental value~\cite{lattice1,latt04}
and others the KN scattering state~\cite{lattice2}.  
The possible existence of such a state has
 raised interesting questions about its structure in connection to 
its narrow width.
A number of phenomenological models have been put forward to explain 
its stability 
such as special flux tube configurations~\cite{nussinov,diamond} 
and diquark formation~\cite{Jaffe}.
Another exotic that might exist and has been proposed in the past~\cite{tetraq}
 is a bound state of
 two quarks and two antiquarks. In this work we calculate the static 
$qq\bar{q}\bar{q}$ and $qqqq\bar{q}$ potentials.
The study of the pentaquark potential is  particularly important  
for our
understanding  of the underlying structure of the 
pentaquark bound states that have been detected in various 
experiments~\cite{experiment,xi}.   

To evaluate the static tetraquark and pentaquark potentials
we construct the Wilson loops
 for  the $qq\bar{q}\bar{q}$ and $qqqq\bar{q}$ systems. 
In the strong coupling approximation minimization of the energy of a system 
of two quarks and two antiquarks 
requires that the two quarks and the two antiquarks are connected 
by the minimal length flux tube.
The flux tubes from each of the quarks and the antiquarks can meet at one 
or at two points.
For the geometries
considered in this work  the length of the flux tube with
two Steiner points is always smaller than the configuration
with the flux tubes meeting at a point
and therefore it corresponds to  the minimal flux tube length.
If the minimal flux tube length is denoted by $L_{\min}$
then  the confining potential in the strong coupling approximation is 
 $\sigma L_{min}$ where
for $\sigma$ we take the string tension extracted 
from the quark-antiquark potential.
Besides comparing the  static tetraquark potential
to the one extracted in the strong coupling approximation
we also compare it  to the sum of two meson potentials. 
Similarly the pentaquark potential is compared to the 
potential extracted in the strong coupling
expansion and to the sum of the baryonic and mesonic potentials.
Minimization of the energy of a pentaquark system with the condition
that  the flux
tubes connecting the quarks meet at three points gives, as minimal length, the
flux tube configuration where the three flux tube junctions
 are  Steiner points.
We do not consider
here geometries resulting in the diamond flux tube arrangement~\cite{diamond} 
for which 
the flux tubes from the four quarks meet at a single point where the antiquark
is located.
Obtaining results for both the tetraquark and pentaquark potentials enables us to
look for 
differences in their  behaviour 
which can reflect different structures 
in the tetraquark and pentaquark systems.  
Such information may lead to  important insight
in our understanding of the structure of the 
$\Theta^+$.

\section{Wilson loops}
\begin{figure}
\epsfxsize=7truecm
\epsfysize=5truecm
\mbox{\epsfbox{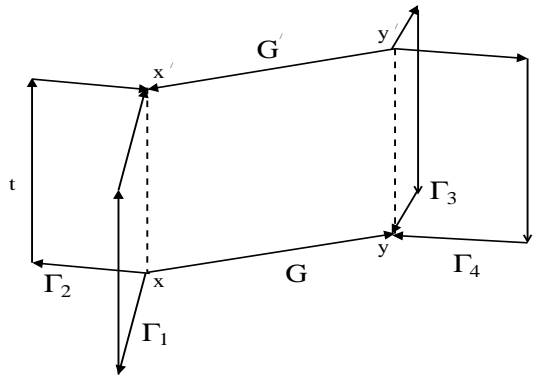}}
\caption{The Wilson loop for the $qq\bar{q}\bar{q}$ system.}
\label{fig:tetraq_loop}
\end{figure}

The SU(3)  Wilson loop for the tetraquark system is constructed by 
creating a gauge invariant
four quark state at time $t=0$, which is annihilated at a later time $t$ as 
shown in Fig.~\ref{fig:tetraq_loop}. 
The two quarks are combined into a colour
$\bar{3}$ and the two antiquarks into a colour $3$ representation of SU(3).
The expression for the tetraquark Wilson loop is given by~\cite{latt04} 
\beq
W_{4q}&=&\frac{1}{12}\epsilon^{abc}\epsilon^{def}\epsilon^{a'b'c'}\epsilon^{d'e'f'}
 U({\bf x,x'},1)^{aa'} U({\bf x,x'},2)^{bb'} \nonumber \\
&\>& \hspace*{-1cm}U_G({\bf x,y})^{cf} U({\bf y',y},3)^{d'd}U({\bf y',y},4)^{e'e} 
      U_{G'}({\bf y',x'})^{f'c'} \quad
\label{4q Wilson}
\eeq
where the two quark lines are created at ${\bf x}$ and
the two antiquark lines at  
${\bf y}$ at time $t=0$ and annihilated at ${\bf x}'$ and ${\bf y}'$ 
at time $t$ respectively.
The staples $U({\bf x},{\bf x'},k)$ involved in the definition of the
Wilson loop are given by
\be
U({\bf x},{\bf x}',k)=P\exp\left[ig\int_{\Gamma_k} dz^{\mu}A_\mu(z)\right]
\ee
where $P$ is the path ordered operator and $\Gamma_k$ denotes the path from ${\bf x}$
to ${\bf x}'$ for quark line $k$ as shown in Fig.~\ref{fig:tetraq_loop}.
The baryonic junction at ${\bf x}$  is joined to the anti-baryonic junction 
at ${\bf y}$  by
\be
U_G({\bf x,y}) = P\exp\left[ig\int_G d{\bf z} . {\bf A}(z)\right]
\ee
at $t=0$. $U_{G'}({\bf y', x'})$ is the corresponding arrow joining the 
anti-baryonic and baryonic junctions
at time $t$.

The tetraquark potential is then extracted
in the standard way from the long time behaviour of the Wilson loop:
\be
V_{4q}=-\lim_{t \rightarrow \infty} \frac{1}{t} \ln<W_{4q}> \quad.
\label{4q pot}
\ee

\begin{figure}
\epsfxsize=7truecm
\epsfysize=6.truecm
\mbox{\epsfbox{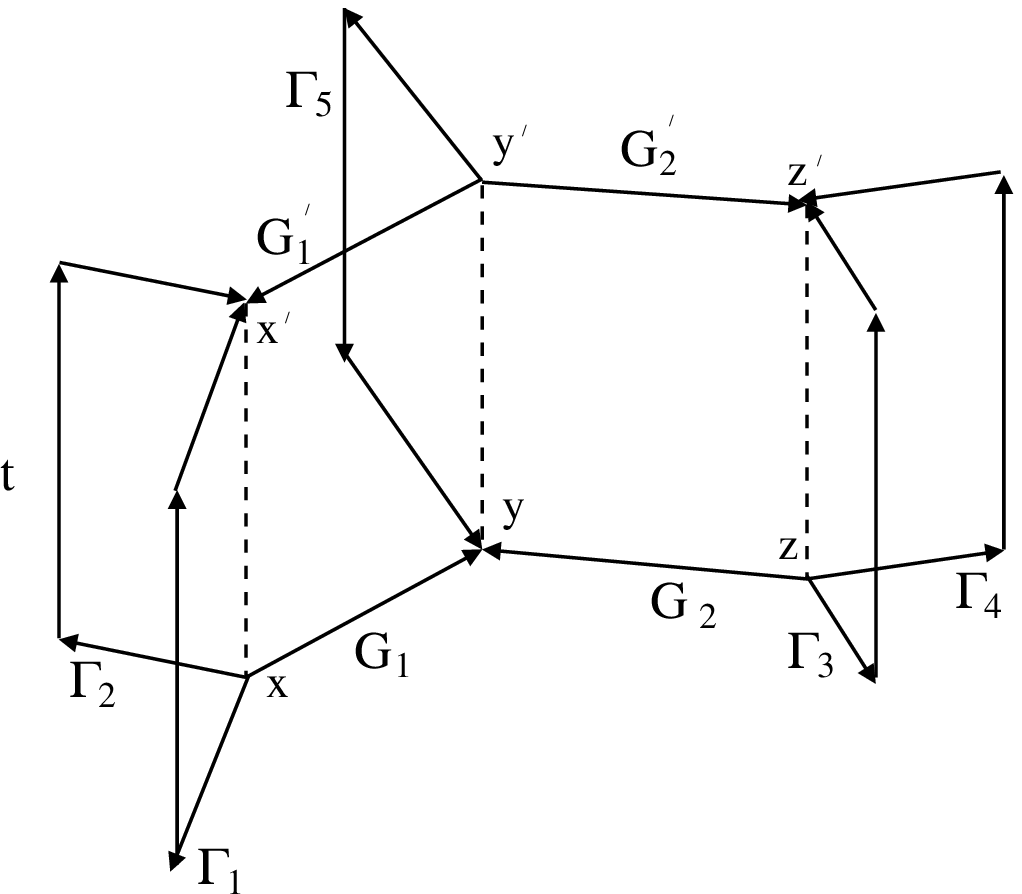}}
\caption{The Wilson loop for the $qqqq\bar{q}$ system.}
\label{fig:pentaq}
\end{figure}

The pentaquark Wilson loop is constructed in a similar way:
 The gauge invariant state evolves now
two baryonic and one anti-baryonic junction. 
The four quarks are grouped into two diquarks each in a colour
$\bar{3}$ representation of SU(3) and the two diquarks
are then combined into a colour singlet with the remaining antiquark.
The Wilson loop is 
shown in Fig.~\ref{fig:pentaq} and it is given by~\cite{latt04}

\beq
W_{5q}&=&\frac{1}{24}\>\>\biggl[\epsilon^{abc}\epsilon^{a'b'c'}
U({\bf x,x'},1)^{aa'} U({\bf x,x'},2)^{bb'}\biggr] \nonumber\\
 &\>&  \hspace*{-1cm}\biggl[\epsilon^{def}\epsilon^{d'e'f'}U({\bf z,z'},3)^{dd'}
 U({\bf z,z'},4)^{ee'}\biggr]U({\bf y',y},5)^{j'j} \nonumber \\
&\>&\epsilon^{ghj}\epsilon^{g'h'j'} U_{G_1}({\bf x,y})^{cg} U_{G_2}({\bf z,y})^{fh}\nonumber \\
&\>& \hspace*{1.5cm}U_{G'_1}({\bf y',x'})^{g'c'} U_{G'_2}({\bf y',z'})^{h'f'} 
\label{5q Wilson}
\eeq
using the same notation as that of Eq.~(\ref{4q Wilson}).

\section{Lattice techniques}
A number of improvements are needed in order to reduce the noise in the measurement of the Wilson loops.
Since the tetraquark and pentaquark potentials are larger than the baryonic potential they are harder
to compute and noise reduction methods are essential.   
We describe briefly the techniques that we use in order to reduce noise
and extract reliably the ground state:

\noindent
1. We use the multi-hit procedure~\cite{multihit} replacing 
the temporal links  by
their average value
\be
U_4(x) \rightarrow \bar{U}_4(x) =\frac{\int dU \> U_4(n) \>e^{\beta S_4(U)}}
{\int dU  \>e^{\beta S_4(U)}}
\label{multihit}
\ee
with $S_4(U)=\frac{1}{N} {\rm Tr}(U_4(n)F^\dagger(n))$ and $F(n)$ the
staple attached to the time
link that is being integrated over. The integration over the links in Eq.~(\ref{multihit}) 
is carried out analytically~\cite{onelink}. 
It has been
shown in SU(2)~\cite{Bali} that replacing the time
links by their average value in this fashion reduces
the error on large Wilson loops of the order of ten.
The factor found in ref.~\cite{Bali} is $x^{2t}\sim 0.889^{2t}$ where
$t$ is the time extent of the Wilson loop.
For the  tetraquark and pentaquark Wilson loops the reduction factor will be
 $x^{4t}$ and $x^{5t}$ respectively giving an even larger noise reduction for the large loops.

\noindent
2. To maximize the overlap of the trial state
with the four or five  quark ground state  we use APE smearing of the
spatial links~\cite{APE}.  Each spatial link is replaced by
a fat link by acting on it with the smearing operator ${\cal{S}}$
defined by
\be
{\cal{S}}{U}_j(x) = {\cal{P}}\Biggl(U_j(x) + \alpha\sum_{k\neq j}\left[
    U_k(x)U_j(x+a\hat{k})U_k^\dagger(x+a\hat{j})\right]\Biggl) ,
\label{APE snearing}
\ee
where ${\cal{P}}$ denotes  projection onto SU(3). This is iterated
$n$ times.
We consider $M$ different levels of smearing and construct an $M\times M$
correlation matrix of Wilson loops. We take $\alpha=1/2$, $M=3$
 and successive number of smearings $n_1=0,n_2=15$ and  $n_3=30$.

The correlation matrices $C(t)$ for the various Wilson loops are
analyzed using a variational method~\cite{variational}. We analyze the
results in 
two different ways:

In the first variant, which we refer to as variational method 1,
 we  solve the generalized eigenvalue problem
\be
C(t)v_k(t) = {\lambda}_k(t)C(t_0)v_k(t) \hspace*{0.5cm}   k=1,...,M  \quad,   
\ee
taking $t_0/a=1$ and extract the potential levels via
\be
aV_k=\lim_{t \rightarrow \infty} -\biggl[\ln
\left(\frac{\lambda_k(t)}{\lambda_k(t-1)}\right)\biggr]
\label{ratio of eigen}
\ee
by fitting to a constant in the range where $aV_k$ becomes time independent
(plateau region).

In the second variant that we call variational method 2 we first
solve the eigenvalue equation
\be
C(t_0)v_k(t_0) = {\lambda}_k(t_0)v_k(t_0)    
\ee
and project the Wilson correlation matrices to the space corresponding to the N largest
eigenvalues
\be
C^N_{ij}(t) =(v_i(t_0), C(t) v_j(t_0))  
\ee
with $N\le M$. We then solve the generalized eigenvalue equation
\be
C^N(t)v_k(t) = {\lambda}_k(t)C^N(t_0)v_k(t)
\ee
in the truncated space.

In addition the potential is extracted by considering the Wilson loop with the largest number of smearings
and fit to  the ratio  
\be
V=\lim_{t\rightarrow \infty} -\ln[W(t)/W(t-1)]
\label{smeared wilson}
\ee
in the plateau region.

The noise reduction techniques described above 
were shown to yield accurate enough 
results for the baryonic potential for inter-quark distances of ${\cal O}(1.5)$~fm
\cite{AFT}.

\section{Results}

All the computations were carried out on a lattice of size $16^3\times 32$
at $\beta$ values $5.8$ and $6.0$ using 200 and 220 configurations respectively
available at the NERSC archive~\cite{connection}. 

 We compute 
the static $q\bar{q}$ potential  
 on the same configurations 
used for the evaluation of the tetraquark and 
pentaquark potentials. Fitting the $q\bar{q}$ potential to the Cornell Ansatz
\be
V_{q\bar{q}}(r)=V_0 -\frac{\alpha}{r} + \sigma r
\label{qqbar}
\ee
we extract the parameters $V_0$, $\alpha$ and string tension $\sigma$.
We give the values obtained in Table~\ref{table:parameters}.
Using the value of $a^2\sigma$ given in  Table~\ref{table:parameters}
and the physical value of $\sqrt{\sigma}=440$~MeV known from Regge theory 
enables us to fix the lattice spacing $a$. We obtain
$a=0.10$~fm at $\beta=6.0$ and $a=0.15$~fm at $\beta=5.8$.
Since in this work we  compare the tetraquark potential to 
the sum of two meson potentials and the pentaquark potential
to the sum of the corresponding
mesonic and baryonic potentials we need
to compute all the potentials with the quarks at the same
locations. For the mesonic potential needed in the comparison
of the tetraquark system we
use lattice data avoiding any parametrizations.
For the baryonic potential needed in the comparison  of
the pentaquark potential we do not have lattice data and therefore
we need a parametrization. Two possible Ans\"atze have been discussed in
the literature for the baryonic potential,
the Y-Ansatz~\cite{CKP} 
and the 
 sum of two-body potentials or $\Delta$-Ansatz.  
The latter was derived using  center vortices~\cite{Cornwall}
but with an  erroneous assumption, which after correction was shown to
support the Y-Ansatz~\cite{Cornwall2}. 
Nevertheless within lattice QCD one can  check
the two Ans\"atze and determine whether the  confining potential
is closer to a sum of two-body potentials or to the Y-Ansatz as a function
of the distance between the quarks. The results of the two Ans\"atze
differ at the most by 15\% and have been
compared to lattice results 
 in refs.~\cite{Sommer,Bali2,AFT,AFJ,japan,map}.
The general consensus from these studies is that  at large distances
the baryonic potential  approaches the
Y-Ansatz. Therefore here  it suffices
to use the Y-Ansatz to parametrize the
baryonic potential. 
As in ref.~\cite{AFT} we use
 the values for $V_0$, $\alpha$
and $\sigma$ extracted from  fitting  the  $q\bar{q}$ potential via
Eq.~(\ref{qqbar}) and given in Table~\ref{table:parameters}.
This means that there are no adjustable parameters in the comparison
of the pentaquark potential to the sum of the mesonic and baryonic potential. 
In addition both the tetraquark and pentaquark potentials are compared to  
 the potential derived 
in the strong coupling expansion.
We use the parametrization given by
\be 
{V_{\rm min}^{4q} \choose V_{\rm min}^{5q}}=
{2 \choose 5/2}  V_0-n_q\sum_{i> j}\frac{\alpha}
{|{\bf r}_i-{\bf r}_j|}+\sigma {L_{\rm min}^{4q} \choose L_{\rm min}^{5q}}
 \quad,
\label{Vmin}
\ee
where again for $V_0, \alpha$ and $\sigma$ we use the values given
in Table~\ref{table:parameters}.
The factor $n_q$ in front of the Coulomb term  is one when a quark 
interacts with 
an antiquark 
and one half when the interaction is
  between quarks or between antiquarks as obtained 
from the one-gluon exchange approximation.

\begin{table}
\caption{The parameters of the static $q\bar{q}$ potential in lattice
units at $\beta=5.8$ and $\beta=6.0$.}
\label{table:parameters}
\begin{tabular}{|r|r|r|r|}
\hline
\multicolumn{1}{|c|}{$\beta $ } &
\multicolumn{1}{ c|}{$aV_0$ } &
\multicolumn{1}{ c|}{$\alpha$ } &
\multicolumn{1}{ c|}{$a^2\sigma $ } \\
\hline
\hline
6.0 &  0.637(5)& 0.255(5) & 0.050(1) \\
\hline
5.8 & 0.636(11)& 0.248(11)& 0.105(2) \\
\hline
\end{tabular}
\end{table}

\begin{figure}[h]
\epsfxsize=7truecm
\epsfysize=4.truecm
\mbox{\epsfbox{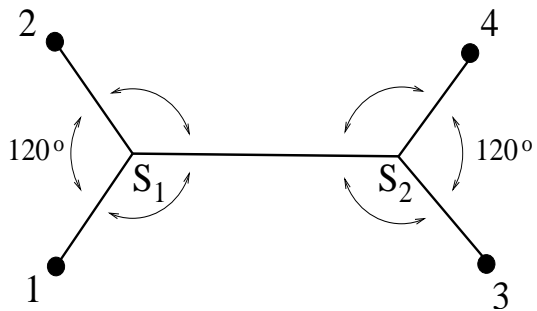}}
\caption{ The tetraquark 
 minimal length. The angles at the junctions
are  120$^0$.}
\label{fig:Lmin tetra}
\end{figure}

\begin{figure}[h]
\epsfxsize=6.5truecm
\epsfysize=4.5truecm
\mbox{\epsfbox{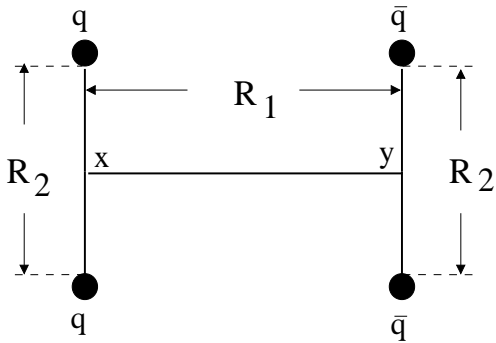}}
\caption{The geometry used for the computation of the tetraquark 
static potential
 where we have
taken the baryonic junction at  the origin and the anti-baryonic
junction at ${\bf y}=(R_1,0,0)$. The quarks are located
at positions ${\bf r}_1=(0,R_2/2,0),{\bf r}_2=(0,-R_2/2,0)$ and the
antiquarks at ${\bf r}_3=(R_1,R_2/2,0)$ and ${\bf r}_4=(R_1,-R_2/2,0)$.}
\label{fig:tetraq_geom}
\end{figure}

\begin{figure}[h]
\epsfxsize=7truecm
\epsfysize=5.5truecm
\mbox{\epsfbox{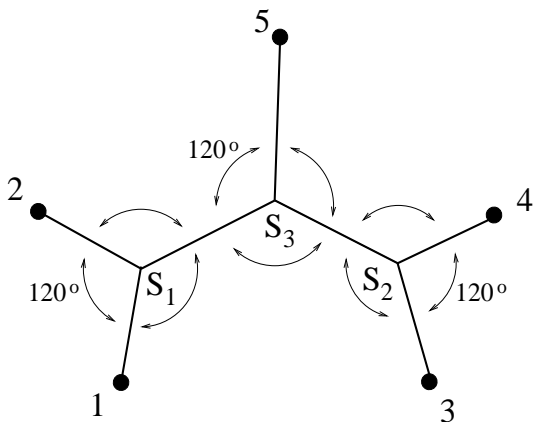}}
\caption{ The pentaquark 
 minimal length. The angles at the junctions
are all 120$^0$.}
\label{fig:Lmin penta}
\end{figure}

\begin{figure}[h]
\epsfxsize=7.5truecm
\epsfysize=10truecm
\mbox{\epsfbox{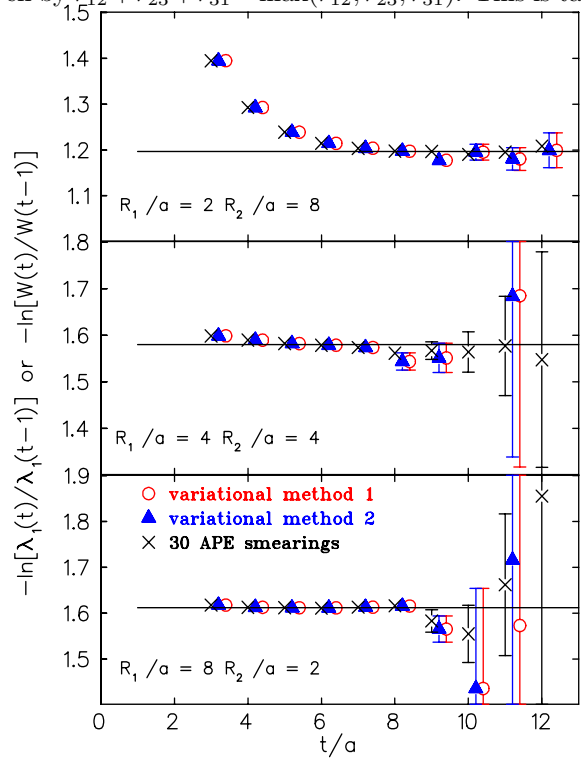}}
\caption{Plateau values for the tetraquark potential. 
Top for the case where  $R_1/a= 2$ and $R_2/a=8$; middle for $R_1/a= 4$ 
and $R_2/a=4$;
bottom for $R_1/a= 8$ and $R_2/a=2$. The data obtained using
variational method 1 are denoted by the  circles,
 using variational method 2 by the filled triangles and 
using Eq.~(\ref{smeared wilson}) by the crosses. The circles and
 filled triangles are shifted horizontally for clarity.}
\label{fig:tetraq_plat}
\end{figure}

 For a general  tetraquark configuration the minimal length having
two Steiner points is  shown in Fig.~\ref{fig:Lmin tetra} and, in general,
is non-planar. For the planar geometry considered in this work
the minimal length can be easily computed and it is given by
\be L_{\rm min}^{4q}=\min \biggl(R_1+\sqrt{3}R_2, R_2+\sqrt{3}R_1 \biggr)\quad,
\label{Lmin 4q}
\ee
where the distances $R_1$ and $R_2$ are defined in
 Fig.~\ref{fig:tetraq_geom}.
For a general  pentaquark configuration   the minimal length
with three junctions is  non planar and has the geometry  shown 
in Fig.~\ref{fig:Lmin penta} with  the three junctions being Steiner points.
We compute these
 three Steiner points numerically
as follows: We start
by an initial guess for the Steiner point $S_3$. Having
an initial guess for $S_3$ enables us to evaluate the two other Steiner
points  analytically in terms of $S_3$ following ref.~\cite{CKP}:
\beq
&\>&|{\bf r}_i-{\bf r}_{S_1}|=\frac{C^2-r_{jk}^2}{B} \hspace*{1cm} (i,j,k 
\hspace*{0.5cm}{\rm cyclic}) \nonumber \\  
C^2 & = &\frac{1}{2}\xi + \bigl(\frac{1}{3} \eta -\frac{1}{12} \xi^2\bigr)^{1/2} \hspace*{0.6cm}
 B^2 = 3C^2-\xi \nonumber \\
\xi &= &r_{12}^2+r_{23}^2+r_{31}^2 \nonumber \\
\eta &=&r_{12}^2 r_{23}^2+r_{23}^2 r_{31}^2+r_{31}^2 r_{12}^2 \quad,
\label{Steiner point}
\eeq
where $r_{12}=|{\bf r}_1-{\bf r}_2|, \> r_{23}=|{\bf r}_2-{\bf r}_{S_3}|$,
and $r_{31}=|{\bf r}_{S_3}-{\bf r}_1|$.
The minimal length joining the three points ${\bf r}_1$,  ${\bf r}_2$
and  ${\bf r}_{S_3}$ is equal to $B$. If any interior 
angle of the triangle ${\bf r}_1{\bf r}_2 {\bf r_{S_3}}$ is
greater than $120^0$
then the minimal length is given by 
$r_{12}+r_{23}+ r_{31}-{\rm max}(r_{12},r_{23}, r_{31})$.
 This is taken into account in our analytic
determination of the Steiner points.
 An analogous
expression holds for $S_2$.
 Having
a value for $S_1$ and $S_2$ and knowing ${\bf r}_5$ 
we can evaluate analytically the new value
for $S_3$. This procedure is iterated until convergence is reached after
a small number (of order 10) of iterations.
We use this method to compute $L_{\rm min}^{5q}$
for the geometry studied in this work. This approach can also be applied to
find the minimal length for a general tetraquark geometry.

\begin{figure}[h]
\epsfxsize=7.5truecm
\epsfysize=5.5truecm
\mbox{\epsfbox{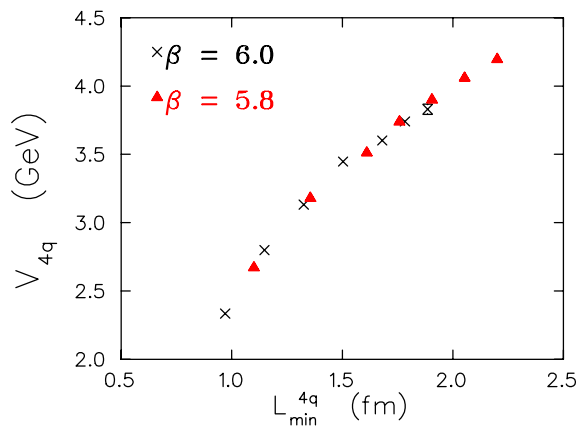}}
\caption{ The tetraquark static potential at $\beta=6.0$ (crosses)
and $\beta=5.8$ (filled triangles) 
for $R_2 = 0.6$~ fm as a function 
of the minimal length in physical units. We have applied a constant shift to 
the data at $\beta=5.8$. The jackknife errors are smaller than the  size of the symbols.}
\label{fig:tetraq_scaling}
\end{figure}

\begin{widetext}

\begin{table}[h]
\caption{Determination of plateau values
for the tetraquark  potential
at $\beta=6.0$. The first column gives the diquark distance, $R_1$, and
for each of the internal diquark distances, $R_2$, we give, the
the initial and final time  $t_i/a$ and $t_f/a$ 
used in fitting the plateau,  the value of the potential
 and  $\tilde{\chi}^2=\chi^2/{\rm d.o.f}$.  We denote with an asterisk the value
that we choose. 
All quantities are given
in lattice units.}
\label{table:tetraq_fit_range}
\begin{tabular}{|c *{2}{|r} |l |c| *{2}{|r} |l |c| *{2}{|r} |l |c| *{2}{|r} |l |c|}
\cline{1-17}
\multicolumn{17}{|c|}{Tetraquark potential} \\ \hline
\multicolumn{1}{|c|} { $\,R_1/a  \,$ }    &
\multicolumn{4}{|c||}{ $\,R_2/a=2\,$ }    &
\multicolumn{4}{|c||}{ $\,R_2/a=4\,$ }    &
\multicolumn{4}{|c||}{ $\,R_2/a=6\,$ }    &
\multicolumn{4}{|c|} { $\,R_2/a=8\,$ }    \\
\cline{1-17}
\multicolumn{1}{|c  }{                 }      & 
\multicolumn{1}{|c  }{$t_{i}/a$}              & 
\multicolumn{1}{|c  }{$t_{f}/a$}              & 
\multicolumn{1}{|c  }{$a V_{4q}$}             & 
\multicolumn{1}{|c||}{$\tilde{\chi}^2$ }      & 
\multicolumn{1}{|c  }{$t_{i}/a$}              & 
\multicolumn{1}{|c  }{$t_{f}/a$}              & 
\multicolumn{1}{|c  }{$a V_{4q}$}             & 
\multicolumn{1}{|c||}{$\tilde{\chi}^2$ }      & 
\multicolumn{1}{|c  }{$t_{i}/a$}              & 
\multicolumn{1}{|c  }{$t_{f}/a$}              &
\multicolumn{1}{|c  }{$a V_{4q}$}             &
\multicolumn{1}{|c||}{$\tilde{\chi}^2$ }      & 
\multicolumn{1}{|c  }{$t_{i}/a$}              &
\multicolumn{1}{|c  }{$t_{f}/a$}              &
\multicolumn{1}{|c  }{$a V_{4q}$}             &
\multicolumn{1}{|c| }{$\tilde{\chi}^2$ }      \\
\cline{2-17}
 &$ 8$&$12$&$1.195(1)  $&$13 $&$ 8$&$12$&$1.219(1)  $&$32 $&$ 8$&$12$&$1.201(1)  $&$5.7$&$ 7$&$12$&$1.200(1)  $&$5.4$\\
2&$10$&$12$&$1.190(1)^*$&$1.2$&$10$&$12$&$1.203(2)^*$&$0.6$&$ 9$&$12$&$1.196(1)^*$&$0.8$&$ 8$&$12$&$1.197(1)^*$&$1.4$\\
 &$11$&$12$&$1.190(2)  $&$0.6$&$11$&$12$&$1.200(3)  $&$0.1$&$11$&$12$&$1.191(4)  $&$0.5$&$ 9$&$12$&$1.195(2)  $&$1.6$\\
\hline
 &$ 5$&$12$&$1.318(1)  $&$1.2$&$ 7$&$12$&$1.450(2)   $&$3.7$&$ 6$&$12$&$1.473(2)  $&$16$&$ 7$&$12$&$1.428(4)  $&$2.4 $\\
 3&$ 6$&$12$&$1.318(1)^*$&$0.8$&$ 8$&$12$&$1.441(4)^*$&$1.2$&$ 8$&$12$&$1.435(6)^*$&$0.8$&$ 8$&$12$&$1.409(7)^*$&$0.2 $\\
 &$ 8$&$12$&$1.317(1)  $&$0.9$&$ 9$&$12$&$1.430(7)   $&$0.1$&$10$&$12$&$1.435(24) $&$0.4$&$ 9$&$12$&$1.407(15) $&$0.3 $\\
\hline
  &$ 4$&$12$&$1.390(1)^*$&$1.2$&$ 4$&$12$&$1.584(1)  $&$5.2$&$ 6$&$12$&$1.646(3)   $&$2.8 $&$ 6$&$12$&$1.645(6)  $&$3.4 $\\
 4&$ 7$&$12$&$1.390(1)  $&$1.6$&$ 6$&$12$&$1.577(2)^*$&$0.8$&$ 8$&$12$&$1.605(16)^*$&$1.4 $&$ 6$&$10$&$1.646(6)^*$&$0.7 $\\
  &$10$&$12$&$1.386(6)  $&$2.6$&$ 9$&$12$&$1.567(20) $&$0.1$&$10$&$12$&$1.560(99)  $&$0.7 $&$ 7$&$10$&$1.629(13) $&$0.3 $\\
\hline
 &$ 4$&$12$&$1.450(1)^*$&$0.6 $&$ 5$&$10$&$1.651(2)^*$&$1.1$&$ 5$&$12$&$1.780(3)  $&$1.4  $&$ 5$&$12$&$1.861(6)   $&$2.4 $\\
5&$ 7$&$12$&$1.451(2)  $&$0.5 $&$ 6$&$10$&$1.648(3)  $&$0.7$&$ 6$&$12$&$1.767(7)^*$&$0.5  $&$ 6$&$12$&$1.823(15)^*$&$0.2 $\\
 &$10$&$12$&$1.447(11) $&$0.8 $&$ 7$&$10$&$1.642(6) $&$0.4 $&$ 8$&$12$&$1.725(42) $&$0.2  $&$ 7$&$12$&$1.826(41)  $&$0.2 $\\
\hline
 &$ 3$&$12$&$1.508(1)^*$&$0.8 $&$ 4$&$12$&$1.716(2)^*$&$0.6$&$ 4$&$12$&$1.881(2)   $&$7.4$&$ 4$&$11$&$2.010(4)   $&$4.3$\\
6&$ 7$&$12$&$1.507(3)  $&$0.1 $&$ 7$&$12$&$1.717(10) $&$0.6$&$ 6$&$12$&$1.846(10)^*$&$1.4$&$ 6$&$11$&$1.948(25)^*$&$0.8$\\
 &$10$&$12$&$1.503(21) $&$0.1 $&$ 9$&$12$&$1.656(64)$&$0.2$&$10$&$12$&$1.278(808)  $&$0.1$&$ 7$&$11$&$2.043(86)  $&$0.7$\\
\hline
  &$ 4$&$12$&$1.560(2)^*$&$0.5$&$ 3$&$ 7$&$1.773(2)  $&$4.0$&$ 4$&$ 9$&$1.933(4)   $&$2.2$&$ 4$&$12$&$2.088(6)   $&$2.0$\\
 7&$ 7$&$12$&$1.560(4)  $&$0.7$&$ 5$&$ 7$&$1.766(3)^*$&$0.7$&$ 6$&$ 9$&$1.918(14)^*$&$0.6$&$ 6$&$12$&$2.015(36)^*$&$0.5$\\
  &$10$&$12$&$1.563(36) $&$1.4$&$ 5$&$ 9$&$1.765(3)  $&$5.7$&$ 7$&$ 9$&$1.909(41)  $&$0.9$&$ 7$&$12$&$1.906(107)  $&$0.4$\\

\hline
 &$ 4$&$12$&$1.612(2)^*$&$0.4$&$ 6$&$12$&$1.804(7)^*$&$1.1$&$ 4$&$10$&$1.987(4)   $&$1.7$&$ 6$&$12$&$2.080(53)  $&$1.9$\\
8&$ 7$&$12$&$1.612(6)  $&$0.6$&$ 7$&$12$&$1.778(19) $&$0.9$&$ 6$&$10$&$1.963(19)^*$&$0.2$&$ 6$&$ 8$&$2.097(54)^*$&$0.9$\\
 &$10$&$12$&$1.574(67) $&$0.4$&$ 9$&$12$&$1.813(144)$&$0.8$&$ 8$&$10$&$1.941(20)  $&$0.3$&$ 7$&$ 8$&$1.916(109) $&$0.1$\\
\hline
\end{tabular}
\end{table}

\end{widetext}

\begin{figure}[h]
\epsfxsize=7.5truecm
\epsfysize=9.truecm
\mbox{\epsfbox{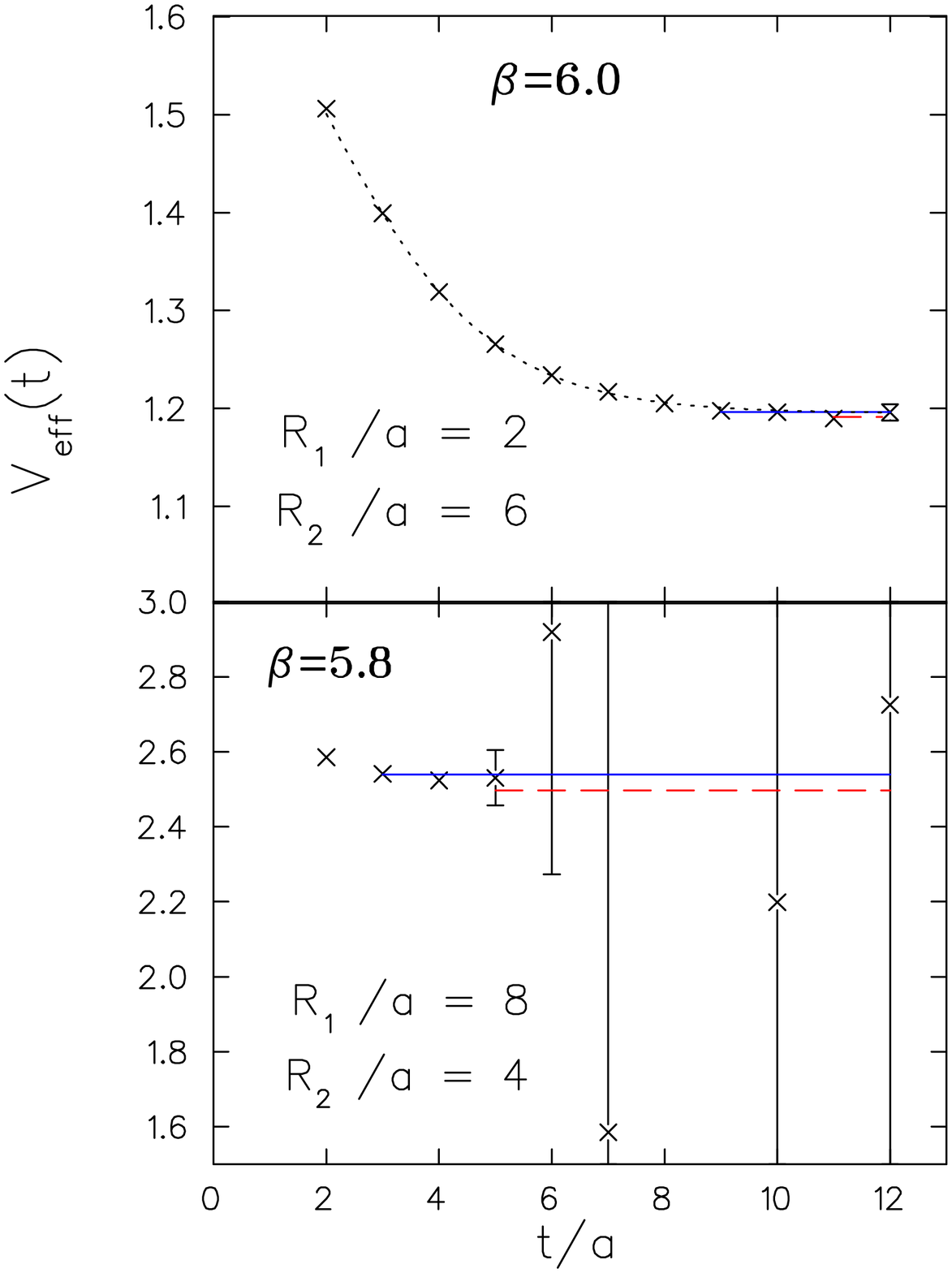}}
\caption{ The ratio $-\log\frac {W(t)}{W(t-1)}$ for
the  tetraquark static potential versus time in lattice units. 
Upper graph  for $R_1=0.2$~fm and $R_2=0.6$~fm at $\beta=6.0$. Lower
graph for $R_1=1.2$~fm and $R_2/a=0.6$~fm at $\beta=5.8$.
Horizontal lines show the value obtained from fitting the lattice
data to a constant with the 
solid line showing the value that we  choose. 
The initial time used in the fit is shown by
the position where the line begins.
The dotted curve shows the result of fitting to Eq.~\ref{V_eff}.}
\label{fig:tetraq_sys}
\end{figure}

We discuss first the results obtained for the tetraquark system.
As already mentioned  the geometry that we use for the computation of the
 tetraquark potential is planar and  is shown 
in Fig.~\ref{fig:tetraq_geom}.
When $R_2$ is small this configuration  allows diquark formation.
We take the distance between the two quarks and the two antiquarks, 
$R_2$, to be equal. We will refer to $R_2$ as the internal diquark distance.
The potential is computed as a function
of the distance between the two diquarks,
$R_1$, and the internal diquark distance $R_2$. 
In Fig.~\ref{fig:tetraq_plat} we show the
plateaus for representative values of the distances $R_1$ and $R_2$. 
On these figures we display the values extracted
using the two variational methods as well as the one extracted from the  
ratio of Wilson loops with the largest number of APE smearings as given in 
Eq.~(\ref{smeared wilson}).
As can be seen from these figures the two variational methods
 yield consistent results of similar quality.
The ratio of the Wilson loops on the other hand yields better plateaus. 
This is true for all the distances that
we have performed this analysis and we will therefore use this method 
to extract the 
potential. We have
always checked that the variational analysis gave results consistent
with those extracted using Eq.~(\ref{smeared wilson}).    
The errors shown on all our figures are jackknife errors.

\begin{figure}[h]
\epsfxsize=7.5truecm
\epsfysize=5.5truecm
\mbox{\epsfbox{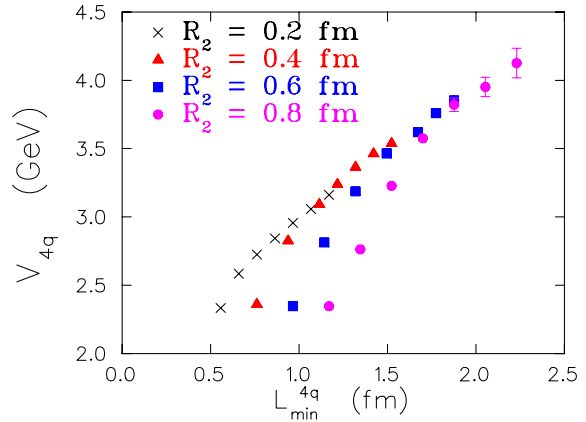}}
\caption{ The tetraquark static potential at $\beta=6.0$ 
versus  the minimal length. Data obtained for $R_2=$0.2, 0.4, 0.6 and
0.8~fm are denoted with crosses, filled triangles, filled squares and 
filled circles respectively.}
\label{fig:tetraq_all}
\end{figure}

\begin{figure}[h]
\epsfxsize=7.5truecm
\epsfysize=5.5truecm
\mbox{\epsfbox{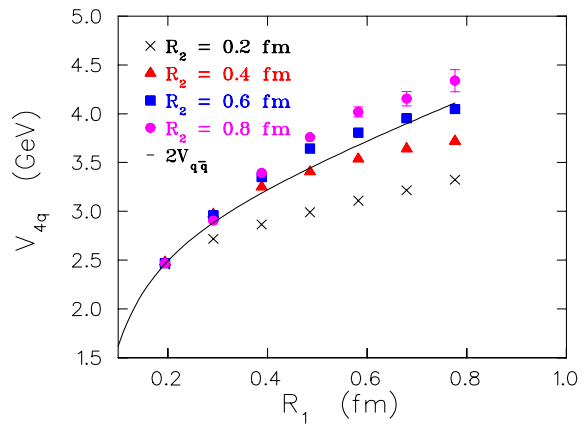}}
\caption{ The tetraquark static potential versus $R_1$ for 
$R_2=$0.2~fm~(crosses), 0.4~fm~(filled triangles), 0.6~fm~(filled squares)
 and 0.8~fm~(filled circles). The
solid line is twice the $q\bar{q}$ potential.}
\label{fig:tetraq}
\end{figure}

\begin{figure}[h]
\begin{minipage}{7.5cm}    
\epsfxsize=7.5truecm
\epsfysize=5.5truecm
\mbox{\epsfbox{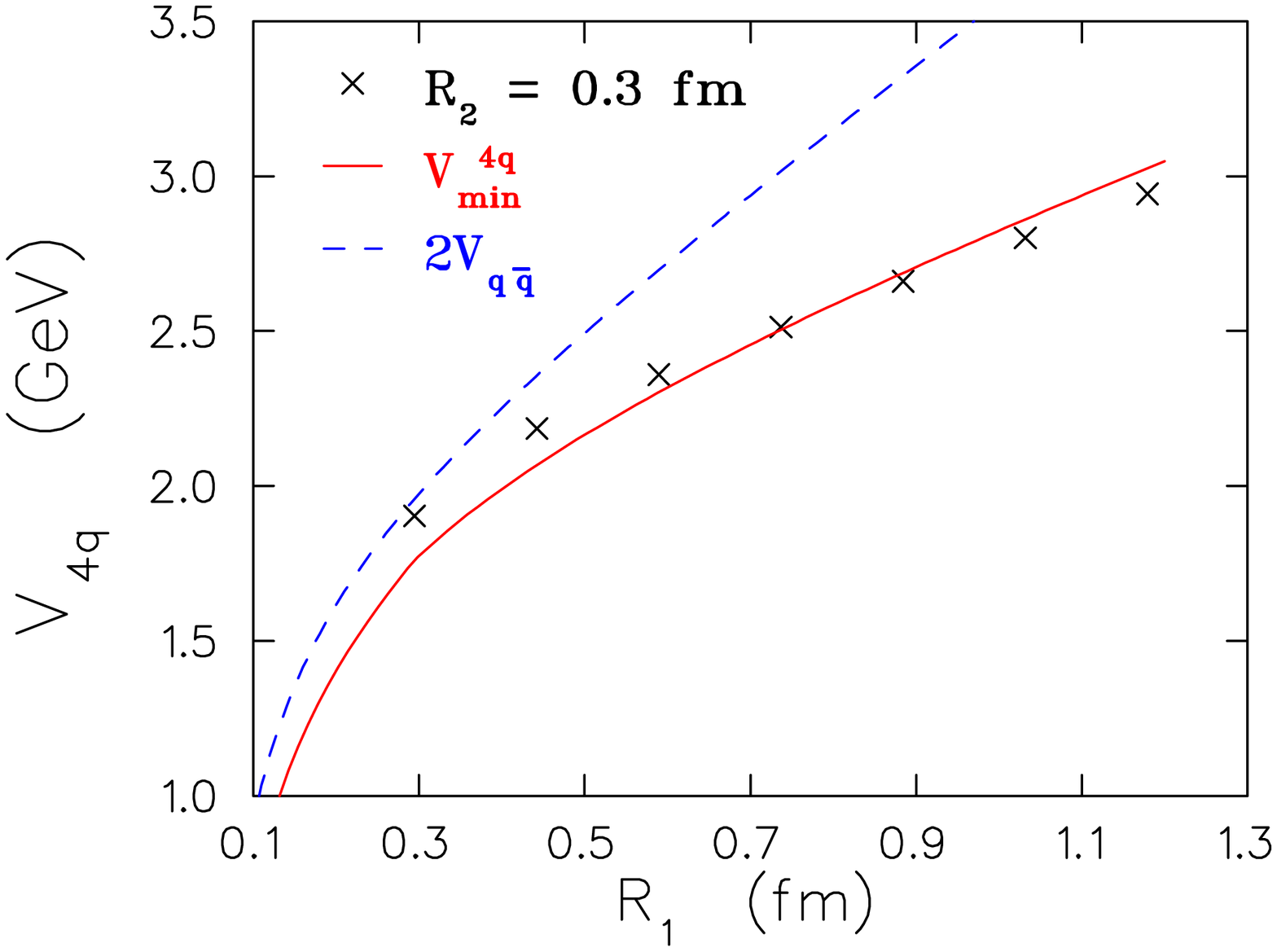}}
\caption{ The tetraquark static potential at $\beta=5.8$ for $R_2 = 0.3$ fm  
compared with $V_{\rm min}^{4q}$ (solid line) and with $2V_{q\bar{q}}(R_1)$ (dashed line).} 
\label{fig:tetraq 03}
\end{minipage}\hfill
\begin{minipage}{7.5cm}
\epsfxsize=7.5truecm
\epsfysize=6truecm
\mbox{\epsfbox{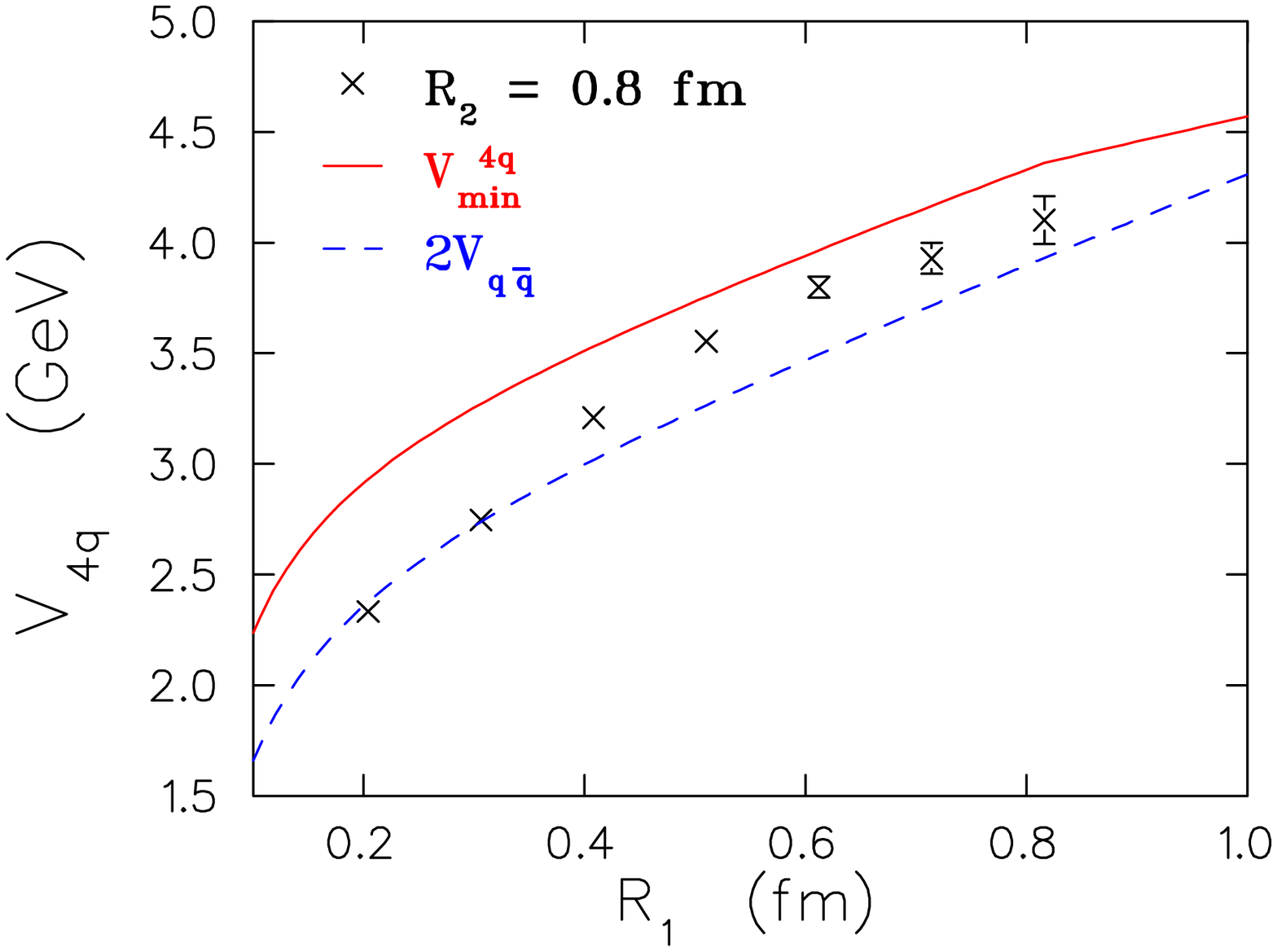}}
\caption{ The tetraquark static potential at $\beta=6.0$ for $R_2 = 0.8$ fm 
compared  with $V_{\rm min}^{4q}$~(solid line) and with
$2V_{q\bar{q}}(R_1)$~(dashed line).}
\label{fig:tetraq 08}
\end{minipage}
\end{figure}

\begin{figure}[h]
\epsfxsize=7truecm
\epsfysize=6truecm
\mbox{\epsfbox{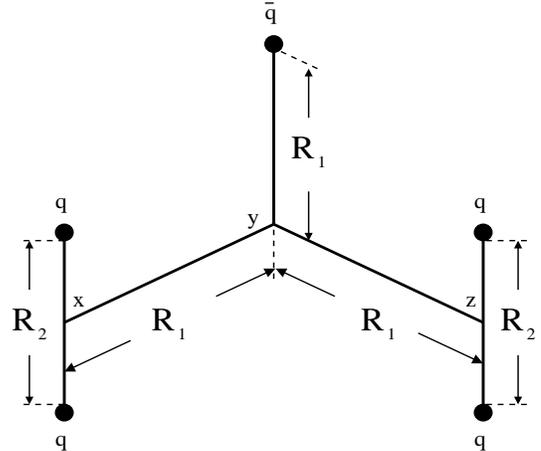}}
\caption{ The geometry used for the computation of the 
pentaquark static potential 
 where we
have taken the baryonic junctions at ${\bf x}=(R_1,0,0)$
and at ${\bf z}=(0,R_1,0)$ and the anti-baryonic junction
at ${\bf y}=(0,0,0)$. The quarks are located
at positions ${\bf r}_1=(R_1,0,R_2/2),{\bf r}_2=(R_1,0,-R_2/2)$, 
${\bf r}_3=(0,R_1,R_2/2)$ and ${\bf r}_4=(0,R_1,-R_2/2)$ and the antiquark at
${\bf r}_5=(0,0,R_1)$.}
\label{fig:pentaq_geom}
\end{figure}

\begin{figure}[h]
\epsfxsize=7.5truecm
\epsfysize=9.5truecm
\mbox{\epsfbox{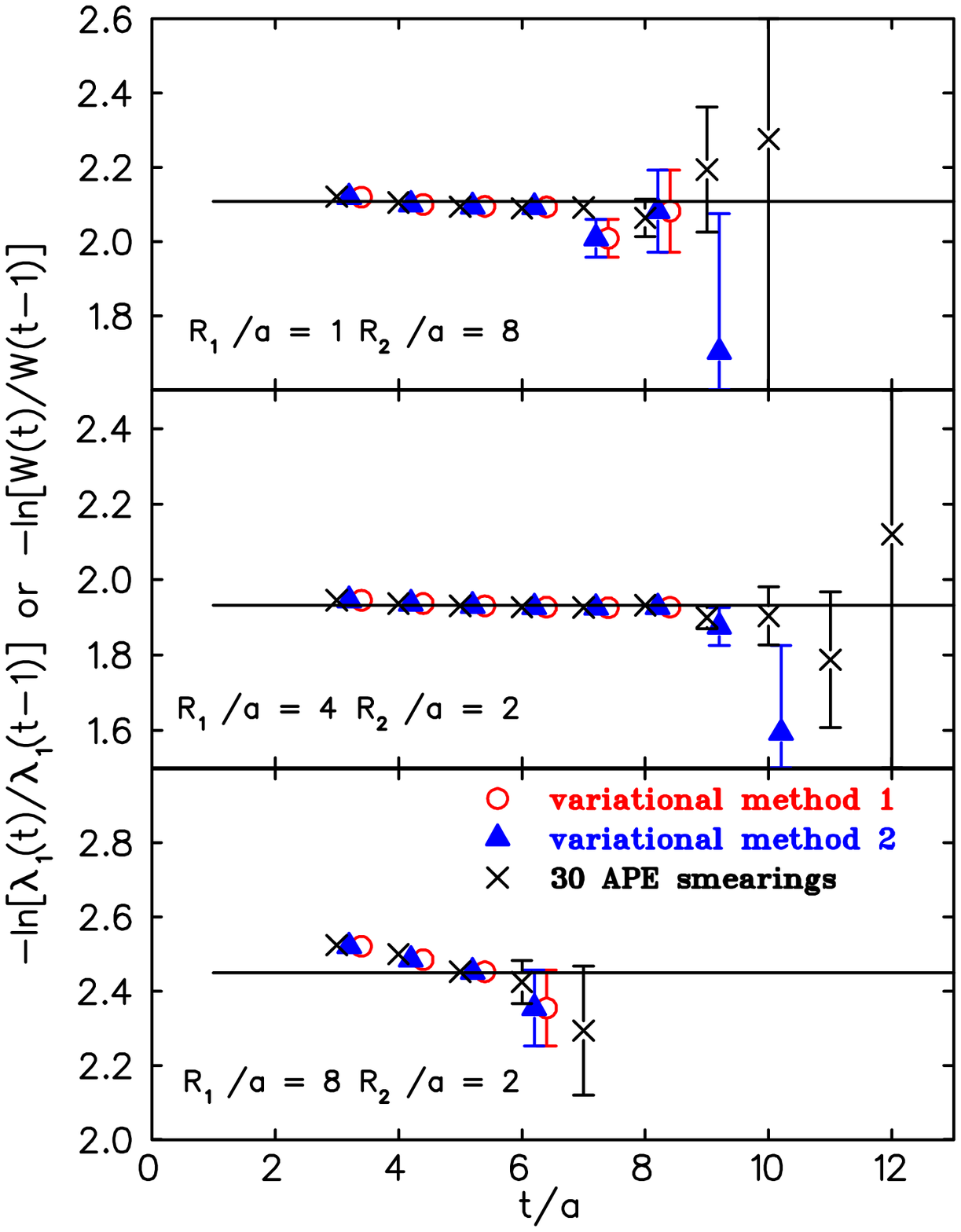}}
\caption{Plateau values for the pentaquark potential. 
Top for the case where $R_1/a= 1$ and $R_2/a=8$; 
middle for $R_1/a= 4$ and $R_2/a=2$;
bottom for $R_1/a= 8$ and $R_2/a=2$. The rest of the notation
is as in Fig.~\ref{fig:tetraq_plat}.}
\label{fig:penta_plat}
\end{figure}

We check scaling by comparing results at $\beta=5.8$ and $\beta=6.0$. 
In Fig.~\ref{fig:tetraq_scaling} we show the static tetraquark potential
for the same physical internal diquark distance $R_2$ as a function of the 
minimal length. We apply a constant
shift to the data at $\beta=5.8$
given  by twice the difference in the constants 
$V_0$ at the two $\beta$ values. 
As can be seen the results, to a good approximation, fall on a universal
line showing reasonable scaling behaviour. 
As in  all our figures
the errors shown on this figure are 
the statistical errors obtained by jackknife analysis.
 To understand why the statistical
errors  are small even at the larger distances we plot in 
Fig.~\ref{fig:tetraq_sys}, for two representative cases, the ratio 
$V_{\rm eff}(t)=-\log\frac{W(t)}{W(t-1)}$, which for large $t$  gives the
ground state potential $V_{4q}$ we are interested in. 
The two cases that we consider are those that have the
smallest and largest values of $L_{\rm min}$
in Fig.~\ref{fig:tetraq_scaling}. The first corresponds to 
$R_1=0.2$~fm 
 at $\beta=6.0$ and the second to $R_1=1.2$~fm at $\beta=5.8$ with $R_2=0.6$~fm in both cases.
To fit to a constant we must search
for a plateau of $V_{\rm eff}(t)$ 
and make sure that changing the initial fitting range does not produce a
 value outside the statistical errors.
 In our analysis we take
the plateau value that gives,
for the earliest initial time range,
 $\tilde{\chi}^2 \equiv\chi^2/{\rm d.o.f.}\stackrel{<}{\sim}1$  
since this criterion ensures a good fit and
gives a result with the smallest statistical error.
When $R_1=0.2$~fm  the results are very accurate and clearly show 
 contributions from excited states with   $V_{\rm eff}(t)$
showing  a plateau only  for $t/a>7$ . Fitting within the plateau range
we extract the values given in Table~\ref{table:tetraq_fit_range} where we
check consistency by  changing the initial fit range.
Using our criterion we choose 
the plateau value with
 $\tilde{\chi}^2=0.8$  
denoted by the asterisk in Table~\ref{table:tetraq_fit_range}.
At these small distances the accuracy of the 
data allows, in addition, a fit that includes the first excited
state. This enables us to extend the time range of the fit and
check that including the first excited state  reproduces
the plateau value obtained for the ground state from fitting to a constant at 
larger times. 
In this case the time-dependence 
of $V_{\rm eff}(t)$ is given by
\be
V_{\rm eff}(t)=V_{4q}-\log\biggl[\frac{1+ce^{-dV t}}{1+ce^{-dV (t-1)}}\biggr]
\label{V_eff}
\ee
where $dV$ is the gap between the ground state  and the first excited
state. Fitting from $t_i/a=3$ we obtain $aV_{4q}=1.195(1)$, which 
agrees with the value extracted by  fitting to a constant 
within the range of the plateau. 
When we consider  the second case at the  largest value of $L_{\rm min}$  
in Fig.~\ref{fig:tetraq_scaling}
we see from Fig.~\ref{fig:tetraq_sys}
that  $V_{\rm eff}(t)$  has very large errors when  $t/a>5$.
This means that the first accurate points used in the fit will
determine the value and error of the extracted constant
and that checking for excited state contributions is no longer possible. 
In this particular
 example fitting to a constant
 from  $t_i/a=3$ gives $\tilde{\chi}^2<1.0$ 
with an error on the extracted value which is small
since it is basically   determined from the two
first accurate points.
Changing the initial fit range to $t_i/a=5$ lowers the plateau
value as shown in the figure but carries an error similar to that of the data point
at $t/a=5$ making it  consistent with the value extracted from the 
fit that used $t_i/a=3$. 
Within our fit criterion we choose the value obtained
from the fit with $t_i/a=3$, which carries a small error
explaining   the small error of the data point 
at the largest value $L_{\rm min}$
in Fig.~\ref{fig:tetraq_scaling}.
We note that for these large interquark distances 
changing the final time range will not affect the value and error of the
extracted potentials
 since the Wilson loops  become very noisy at the larger times. 
In this particular example whether we take the 
upper fit range  to be $t_f/a=7$ or $t_f/a=12$ does not affect
the  value and error resulting from the fit. 
As can be seen from Table~\ref{table:tetraq_fit_range} 
in general we choose the larger  available value of $t_f/a$ 
but for the large distances we reduce the upper fit range
when the 
data points become too noisy.
Since our accurate data points are limited to smaller times the plateau values
that we extract may have  excited state contributions,
 which can not be estimated
since  small shifts in the data points are not visible due
to their large  errors.
However the fact that the data at the 
larger values of $L_{\rm min}$ fall on the same curve as the better determined
points is an indication that this error is reasonably small.
In Table~\ref{table:tetraq_fit_range} we give 
the dependence of the plateau values on the fitting range and our choice
of the value plotted in the figures. The table is done for $\beta=6.0$  
since this is the case that
we discuss in more detail in the manuscript. The analysis at $\beta=5.8$ 
is carried out in the same way.
%
%

\begin{figure}[h]
\epsfxsize=7.5truecm
\epsfysize=5.5truecm
\mbox{\epsfbox{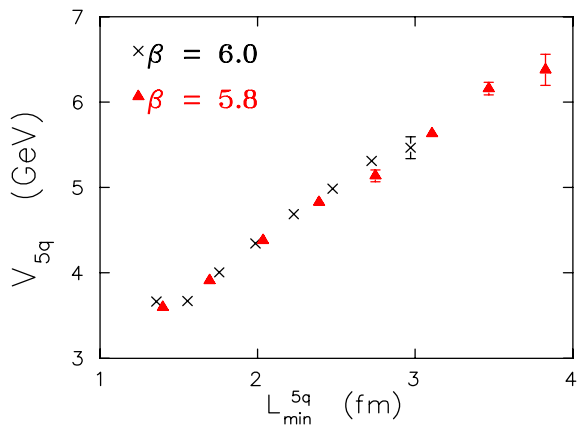}}
\caption{ The pentaquark static potential at $\beta=6.0$~(crosses)
and $\beta=5.8$~(filled triangles) 
for $R_2 = 0.6$~fm as a function of the minimal length. The
data at $\beta=5.8$ are shifted by a constant.} 
\label{fig:pentaq_scaling}
\end{figure}
 
\begin{figure}[h]
\epsfxsize=7.5truecm
\epsfysize=6.truecm
\mbox{\epsfbox{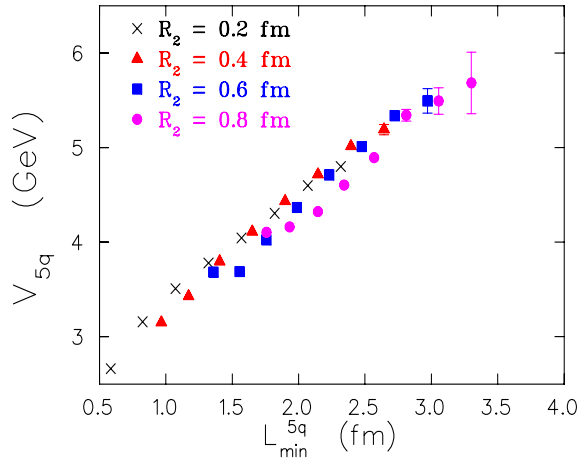}}
\caption{ The pentaquark static potential at $\beta=6.0$ 
versus the minimal length. The notation is the same as that
used in Fig.~\ref{fig:tetraq_all}.}
\label{fig:pentaq_all}
\end{figure}

\begin{figure}[hb]
\epsfxsize=7.5truecm
\epsfysize=5.5truecm
\mbox{\epsfbox{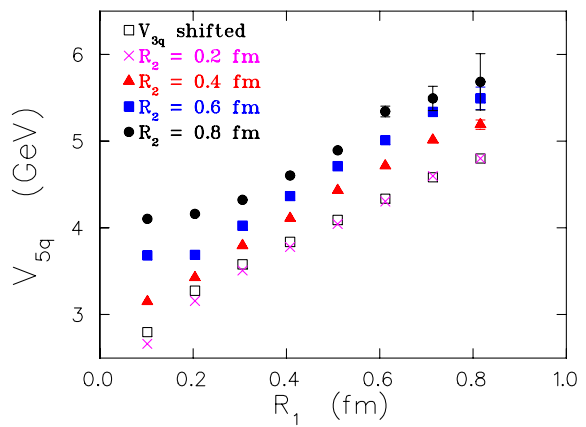}}
\caption{ The pentaquark static potential versus $R_1$ for
$R_2=$0.2~fm~(crosses), 0.4~fm~(filled triangles), 0.6~fm~(filled squares)
 and 0.8~fm~(filled circles). The baryonic potential shifted by 
a constant is shown by the open squares.}
\label{fig:pentaq_pot}
\end{figure}

In Fig.~\ref{fig:tetraq_all} we show the results for the tetraquark potential
at $\beta=6.0$ as a function of the minimal length. 
As  can be seen we have four different
sets of curves for each internal diquark separation $R_2$=0.2, 0.4, 0.6 and 0.8~fm. 
When the distance between the diquarks becomes greater than 
the internal diquark distance
the four sets of curves converge  to the same line. 
This means that for $R_1 > R_2$ the
tetraquark potential becomes approximately
only a function of $L_{\rm min}^{4q}$. On the other hand when $R_1 \le R_2$  and
for $R_1\stackrel{<}{\sim}0.4$~fm
 the potential is independent of $R_2$. 
This behaviour is better seen
in Fig.~\ref{fig:tetraq} where we show the tetraquark potential 
as a function of 
$R_1$ for the four different values of the internal diquark distance 
$R_2$:   
When $R_1=0.2$~fm the potentials coincide for all four different values of $R_2$.
When $R_1=0.3$~fm and, to a good approximation, also when $R_1=0.4$~fm the potentials coincide 
only for $R_2=0.4, 0.6, 0.8$~fm.  
In fact, for $R_1=0.2$~fm and 0.3~fm and $R_2\ge R_1$,
 the tetraquark potential is  given by the
sum of the two meson potentials  as  can be
seen
by the agreement of these results with the line on the 
figure showing  $2V_{q\bar{q}}(R_1)$. This means that for these geometries
 we have 
a system of two mesons rather than a
genuine four quark bound state. 
On the other hand when the distance between the diquarks 
 is 
larger than their internal diquark distance 
 the potential is lower than the sum 
of the two $q\bar{q}$ potentials. 
In  Fig.~\ref{fig:tetraq 03} 
we show lattice data for $\beta=5.8$ for the smallest internal diquark separation,
namely for  $R_2=0.3$~fm, and compare them with $V_{\rm min}^{4q}$
and $2V_{q\bar{q}}(R_1)$.
We have chosen $\beta=5.8$ in order to reach larger diquark separations.  
The tetraquark potential starts as a sum of the two meson potentials
and then crosses over to approach $V_{\rm min}^{4q}$.
Although $V_{\rm min}^{4q}$  has a larger slope as compared to
the slope of the tetraquark confining potential it approximates 
best the lattice data for  values of $R_1$ in the range of about 
one 0.5 to 1~fm.
In Fig.~\ref{fig:tetraq 08}
we perform the same comparison
but  for  internal diquark distance equal to 0.8~fm. 
In this case the tetraquark potential becomes larger than the sum
of the two meson potentials
approaching  $V_{\rm min}^{4q}$ from below. For the two largest
distances 
 the data become too noisy  and better statistics are required to
determine more accurately the long distance behaviour.

We perform a similar analysis for the pentaquark system.
The geometry used for the pentaquark potential is shown 
in Fig.~\ref{fig:pentaq_geom}: We place the
antiquark on the z-axis distance $R_1$ from the origin. 
The two pairs of quarks 
are placed at distances $(R_1,0,\pm R_2/2)$ 
and $(0, R_1,\pm R_2/2)$ so that when $R_2=0$ this configuration reduces 
to the geometry that we used for the baryonic potential~\cite{AFT}.
The quality of the plateaus  is
seen in Fig.~\ref{fig:penta_plat} for
representative values of $R_1$ and $R_2$. Again the
ratio of Wilson loops with the largest number of APE smearings produces
 the best plateaus. 

\begin{widetext}
\begin{table}
\caption{Determination of plateau values for the pentaquark  potential at $\beta=6.0$. The notation is 
the same as that of Table~\ref{table:tetraq_fit_range}.}
\label{table:pentaq_fit_range}
\begin{tabular}{|c *{2}{|r} |l |c| *{2}{|r} |l |c| *{2}{|r} |l |c| *{2}{|r} |l |c|}
\hline
\multicolumn{17}{|c|}{Pentaquark potential} \\ \hline
\multicolumn{1}{|c|} { $\,R_1/a  \,$ }    &
\multicolumn{4}{|c||}{ $\,R_2/a=2\,$ }    &
\multicolumn{4}{|c||}{ $\,R_2/a=4\,$ }    &
\multicolumn{4}{|c||}{ $\,R_2/a=6\,$ }    &
\multicolumn{4}{|c|} { $\,R_2/a=8\,$ }    \\
\cline{1-17}
\multicolumn{1}{|c  }{                 }      & 
\multicolumn{1}{|c  }{$t_{i}/a$}              & 
\multicolumn{1}{|c  }{$t_{f}/a$}              & 
\multicolumn{1}{|c  }{$a V_{5q}$}             & 
\multicolumn{1}{|c||}{$\tilde{\chi}^2$ }      & 
\multicolumn{1}{|c  }{$t_{i}/a$}              & 
\multicolumn{1}{|c  }{$t_{f}/a$}              & 
\multicolumn{1}{|c  }{$a V_{5q}$}             & 
\multicolumn{1}{|c||}{$\tilde{\chi}^2$ }      & 
\multicolumn{1}{|c  }{$t_{i}/a$}              & 
\multicolumn{1}{|c  }{$t_{f}/a$}              &
\multicolumn{1}{|c  }{$a V_{5q}$}             &
\multicolumn{1}{|c||}{$\tilde{\chi}^2$ }      & 
\multicolumn{1}{|c  }{$t_{i}/a$}              &
\multicolumn{1}{|c  }{$t_{f}/a$}              &
\multicolumn{1}{|c  }{$a V_{5q}$}             &
\multicolumn{1}{|c| }{$\tilde{\chi}^2$ }      \\
\cline{2-17}
 &$11$&$12$&$1.369(1)  $&$2.8$ &$ 7$&$11$&$1.616(1)  $&$8.5$&$ 5$&$12$&$1.885(2)   $&$2.9  $&$ 3$&$12$&$2.111(2)   $&$7.5$\\
1&$ 3$&$12$&$1.358(1)^*\footnote[1]{This value was extracted using Eq.~\ref{V_eff}.}  $&$0.2$ &$ 9$&$11$&$1.607(3)^*$&$0.4$&$ 6$&$12$&$1.878(3)^* $&$1.0 $&$ 5$&$12$&$2.093(4)^* $&$0.2$\\
 &$ 5$&$12$&$1.358(2)\footnotemark[1]$&$0.2 $&$10$&$11$&$1.609(5) $&$0.5$&$ 9$&$12$&$1.880(23)  $&$0.9 $&$ 7$&$12$&$2.090(19) $&$0.2$\\
\hline
 &$ 6$&$11$&$1.620(1)  $&$8.6 $&$ 6$&$11$&$1.760(1)  $&$6.3$&$ 6$&$12$&$1.926(3)   $&$3.4 $&$ 5$&$12$&$2.122(4)^* $&$1.0$\\  
2&$ 9$&$11$&$1.611(2)^*$&$0.1 $&$ 8$&$11$&$1.748(4)^*$&$0.9$&$ 8$&$12$&$1.881(13)^*$&$0.6 $&$ 6$&$12$&$2.112(7)   $&$0.7$\\ 
 &$10$&$11$&$1.611(4)  $&$0.2 $&$ 9$&$11$&$1.742(7)  $&$0.8$&$ 9$&$12$&$1.862(32)  $&$0.7 $&$ 8$&$12$&$2.091(60)  $&$0.7$\\
\hline
 &$ 5$&$12$&$1.789(1)^*$&$0.9$&$ 4$&$11$&$1.960(1)  $&$13  $&$ 5$&$12$&$2.068(3)   $&$2.7 $&$ 4$&$10$&$2.228(3)   $&$9.0  $\\  
3&$ 8$&$12$&$1.786(4)  $&$0.2$&$ 7$&$11$&$1.936(5)^*$&$0.1 $&$ 6$&$12$&$2.053(5)^* $&$0.3 $&$ 5$&$10$&$2.205(5)^* $&$1.0  $\\ 
 &$10$&$12$&$1.781(20) $&$0.4$&$ 9$&$11$&$1.928(37) $&$0.1 $&$ 7$&$12$&$2.043(11)  $&$0.2 $&$ 6$&$10$&$2.187(11)  $&$0.4  $\\
\hline
 &$ 5$&$12$&$1.928(2)^*$&$0.5 $&$ 5$&$ 8$&$2.105(3)  $&$1.3 $&$ 3$&$11$&$2.258(3)   $&$18  $&$ 3$&$ 8$&$2.399(3)   $&$25  $\\  
4&$ 8$&$12$&$1.926(12) $&$0.4 $&$ 6$&$ 8$&$2.097(5)^*$&$0.2 $&$ 5$&$11$&$2.227(3)^* $&$1.1 $&$ 5$&$ 8$&$2.348(7)^* $&$0.5 $\\ 
 &$10$&$12$&$1.888(78) $&$0.2 $&$ 7$&$ 8$&$2.105(15) $&$0.1 $&$ 7$&$11$&$2.222(37)  $&$1.3 $&$ 7$&$ 8$&$2.249(96)  $&$0.1 $\\
\hline
 &$ 5$&$11$&$2.063(3)^*$&$0.5$&$ 3$&$ 9$&$2.271(3)  $&$5.7$&$ 4$&$ 9$&$2.403(4)   $&$1.6$&$ 4$&$ 7$&$2.534(6)   $&$4.0 $\\ 
5&$ 7$&$11$&$2.068(11) $&$0.4$&$ 4$&$ 9$&$2.261(3)^*$&$0.6$&$ 4$&$ 7$&$2.403(4)^* $&$1.0$&$ 5$&$ 7$&$2.496(14)^*$&$0.5$\\
 &$ 8$&$11$&$2.069(36) $&$0.3$&$ 7$&$ 9$&$2.277(40) $&$0.5$&$ 5$&$ 7$&$2.391(8)   $&$0.1$&$ 6$&$ 7$&$2.462(62)  $&$0.8$\\
\hline
 &$ 3$&$10$&$2.215(3)  $&$5.6$&$ 4$&$ 8$&$2.405(5)^*$&$1.0$&$ 4$&$10$&$2.556(6)^* $&$1.0$&$ 3$&$ 9$&$2.737(6)   $&$3.9$\\  
6&$ 5$&$10$&$2.196(6)^*$&$0.5$&$ 5$&$ 8$&$2.395(8)  $&$0.6$&$ 6$&$10$&$2.497(55)  $&$0.4$&$ 5$&$ 9$&$2.725(32)^*$&$1.7$\\  
 &$ 7$&$10$&$2.192(29) $&$0.2$&$ 7$&$ 8$&$2.322(98) $&$0.3$&$ 7$&$10$&$2.492(306) $&$0.5$&$ 6$&$ 9$&$2.590(174) $&$0.2$\\ 
\hline
 &$ 4$&$ 8$&$2.345(6)^*$&$1.0$&$ 3$&$ 7$&$2.572(5)  $&$2.6$&$ 4$&$ 9$&$2.722(9)^* $&$0.4$&$ 3$&$ 8$&$2.904(7)    $&$1.9$\\  
7&$ 5$&$ 8$&$2.333(10) $&$0.4$&$ 4$&$ 7$&$2.558(6)^*$&$1.0$&$ 5$&$ 9$&$2.694(29)  $&$0.2$&$ 5$&$ 8$&$2.807(72)^* $&$1.2$\\ 
 &$ 6$&$ 8$&$2.315(24) $&$0.3$&$ 5$&$ 7$&$2.553(16) $&$0.1$&$ 7$&$ 9$&$2.609(148) $&$0.1$&$ 6$&$ 8$&$2.290(279)  $&$0.1$\\
\hline
 &$ 3$&$7$&$2.513(5)   $&$5.7$&$ 3$&$ 9$&$2.724(6)   $&$3.1$&$ 5$&$ 8$&$2.807(66)^*$&$1.8$&$ 5$&$6$&$2.968(165)^*$&$1.1$\\
8&$ 5$&$7$&$2.448(18)^*$&$0.5$&$ 5$&$ 9$&$2.648(28)^*$&$1.1$&$ 6$&$ 8$&$2.266(245) $&$0.2$&    &   &              &     \\
 &$ 6$&$7$&$2.411(55)  $&$0.5$&$ 6$&$ 9$&$2.450(116) $&$0.1$&$ 7$&$ 8$&$2.079(767) $&$0.4$&    &   &              &     \\
\hline
\end{tabular}
\end{table}

\end{widetext}

In Fig.~\ref{fig:pentaq_scaling} we compare the results 
for the pentaquark potential
at $\beta=5.8$ and $6.0$ for the same physical internal diquark 
separation $R_2$. The data at $\beta=5.8$ are shifted by 5/2 the
difference in the values of $V_0$ at the two $\beta$ values.
As in the case of the tetraquark potential, 
the results for the pentaquark potential 
approximately fall on the same curve
indicating good scaling. The comments made for the errors 
on the results obtained at 
$\beta=5.8$ at the larger distances when discussing the scaling of the
tetraquark potential apply also here.
Table~\ref{table:pentaq_fit_range} gives details of our fitting procedure
at $\beta=6.0$ and 
provides
an indication of the systematic errors involved 
 in changing the fitting range.

\begin{figure}[h]
\begin{minipage}{7.5cm}    
\epsfxsize=7.5truecm
\epsfysize=5.5truecm
\mbox{\epsfbox{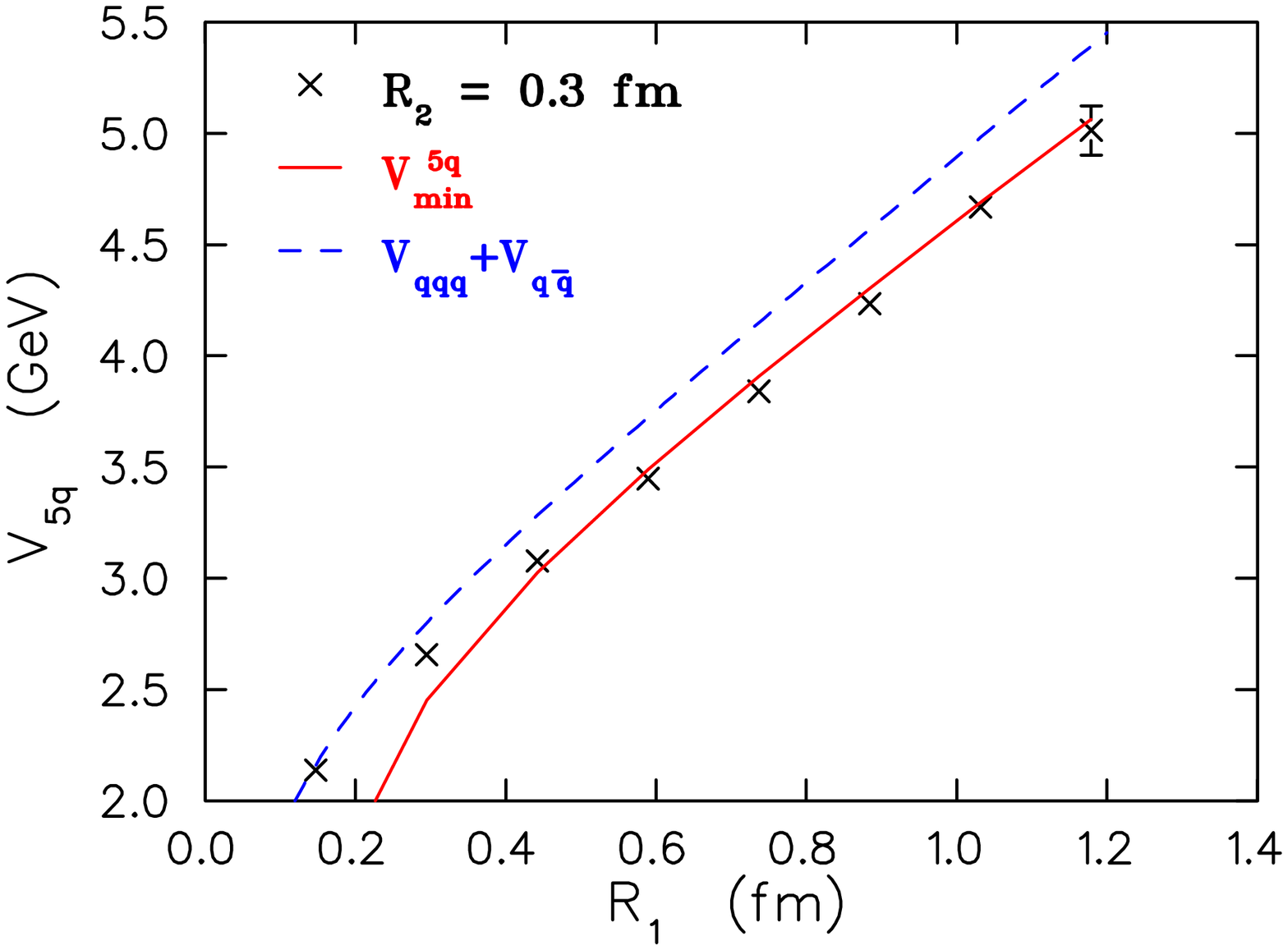}}
\caption{ The pentaquark static potential at $\beta=5.8$ for $R_2 = 0.3$ fm  
compared with $V_{\rm min}^{5q}$ (solid line) and with the sum of the  baryonic 
and  mesonic potentials~(dashed line). }
\label{fig:pentaq 03}
\end{minipage}\hfill
\begin{minipage}{7.5cm}
\epsfxsize=7.5truecm
\epsfysize=5.5truecm
\mbox{\epsfbox{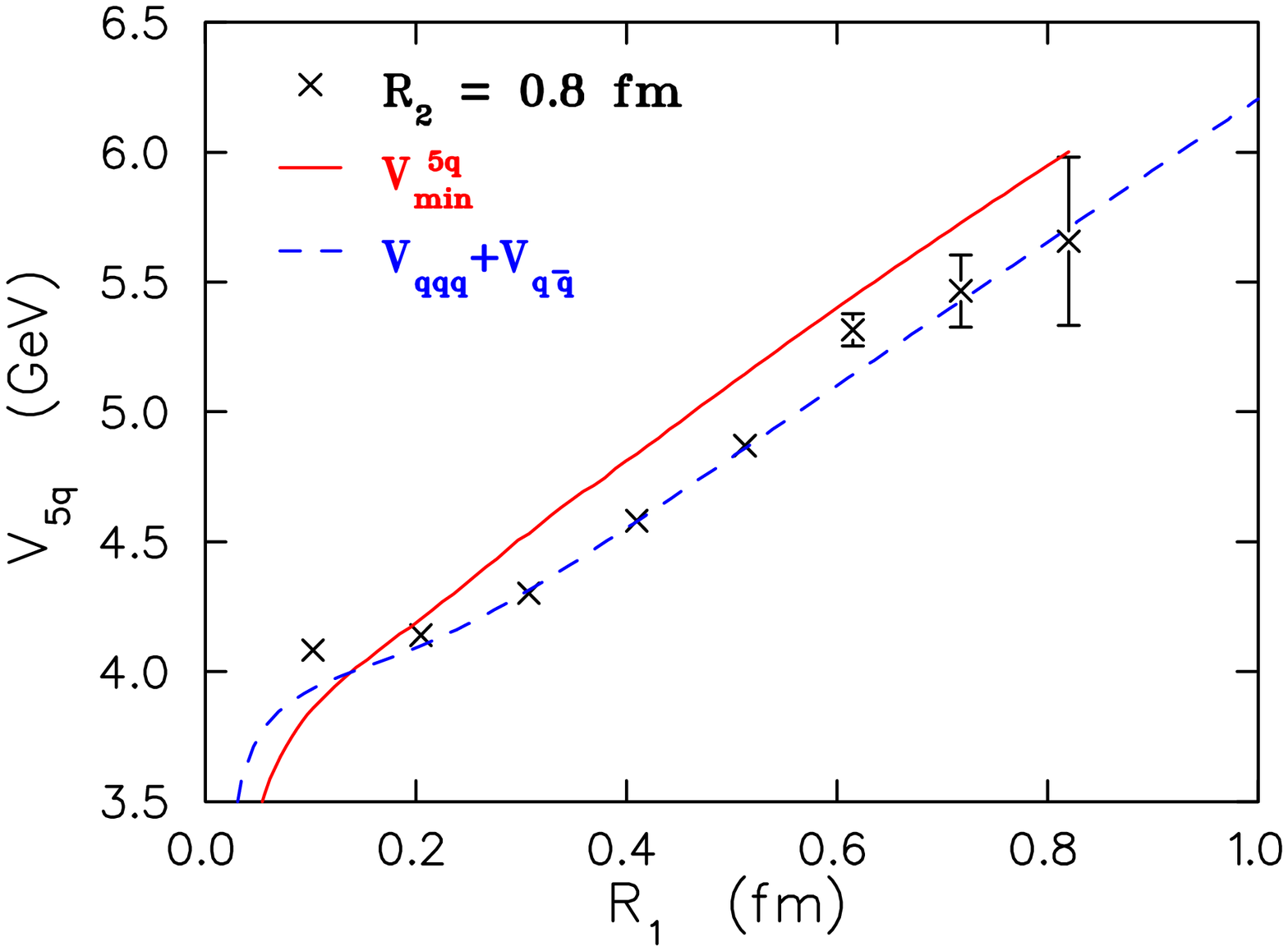}}
\caption{ The pentaquark static potential at $\beta=6.0$ for $R_2 = 0.8$ fm 
compared  with $V_{\rm min}^{5q}$~(solid line)
 and with the sum of the baryonic and mesonic potentials~(dashed line).}
\label{fig:pentaq 08}
\end{minipage}
\end{figure}

In Fig.~\ref{fig:pentaq_all} we plot the pentaquark potential evaluated at 
$\beta=6.0$ versus  $L_{\rm min}^{5q}$.
Although
not as clearly seen as in the case of the tetraquark system
 there are four
sets of curves corresponding to the four different values of $R_2$. For large
values of $R_1$ they converge to the  same curve which again suggests that
the potential asymptotically depends only on the minimal length 
at least to a first approximation.
The four sets
of curves are more clearly seen    
 in Fig.~\ref{fig:pentaq_pot} where
the pentaquark potential is plotted as a function of the diquark 
separation $R_1$ for
the four values of the internal diquark distance $R_2=0.2, 0.4, 0.6$ 
and $0.8$~fm.
On the same figure we also display
 our lattice results for the baryonic potential
shifted by a constant showing that 
the linear dependence of the pentaquark potential for small separations $R_2$
is approximately the same as  that of  the baryonic potential.
It should be noted that, for the geometry used in the  evaluation of the 
pentaquark potential, 
the sum of the baryonic and mesonic potentials depends
on both $R_1$ and $R_2$ and therefore we can no longer draw a universal curve
for all separations $R_2$ as we did for the case of the tetraquark potential.
Instead we compare the pentaquark potential to $V_{\rm min}^{5q}$ and
the sum of the baryonic and mesonic potentials
in Figs.~\ref{fig:pentaq 03} and \ref{fig:pentaq 08} 
for two extreme cases: In Fig.~\ref{fig:pentaq 03} 
we show results for  the smallest possible
internal diquark distance, namely
for $R_2=0.3$~fm, and consider $\beta=5.8$ so that
we can reach larger physical distances in $R_1$.
In this case  the potential 
approaches $V_{\rm min}^{5q}$ for $R_1>0.4$~fm. This means that
when $R_1$ becomes larger than the internal diquark distance  
the $qqqq\bar{q}$ system is well  described by the minimal flux connecting 
the quarks.
This genuine pentaquark state has   
static energy which is lower than the sum of the baryonic 
plus the mesonic potential.
In Fig.~\ref{fig:pentaq 08} 
we show results for  the largest possible internal diquark separation,
namely for $R_2=0.8$~fm, 
and consider $\beta=6.0$ since the results for $\beta=5.8$ become too
noisy to extract any useful information. For this geometry the quarks 
in the diquarks
are always a  distance larger or equal to $R_1$. 
In this
case
the results are well described by 
the sum of the baryonic and  mesonic potentials 
and only for  distances larger than 0.6~fm 
they 
 begin to approach $V_{\rm min}^{5q}$ from below.
More statistics are required  to reduce the errors
in order  to draw a definite conclusion for the large distance
 behaviour in this case.

\section{Conclusions}
We have calculated
the static tetraquark and pentaquark potentials
selecting geometries that are motivated by the diquark picture.
Using multi-hit for the time links and APE smearing for the spatial links
we obtain results for inter-quark distances of the order of 1 fm.
Comparing the results obtained at $\beta=5.8$ and $6.0$ 
we show that the potentials
have reasonable scaling properties. 

The  main conclusions regarding the
tetraquark potential are: 1) When the two diquarks are closer 
than $\sim 0.5$~fm
and with internal diquark distances larger than their separation ($R_2 > R_1$)
then the $qq\bar{q}\bar{q}$ system breaks, as expected, 
into two mesons and its static
potential is approximately given by the sum of the
 two meson potentials.
The geometry used for the tetraquark potential makes it easy to  see when 
the ground state is a two meson state
 since then the static potential  becomes independent of the 
internal diquark distance $R_2$.
2) For distances between diquarks
larger than the internal diquark distance
the static potential is
 approximately described  by $V_{\rm min}^{4q}$, 
which has a confining part proportional to the minimal
length flux tube joining the quarks. 
In the parametrization of $V_{\rm min}^{4q}$ 
we used
values for $V_0, \alpha$ and $\sigma$ extracted from the  
static $q\bar{q}$ potential and therefore there are no
adjustable parameters. 
The linear dependence of the tetraquark potential would require a 
smaller value of $\sigma$
as compared to that extracted from the $q\bar{q}$ potential. 
For internal diquark separations between about 0.5 and 1~fm the
potential is larger than the sum of the two meson potentials and
approaches $V_{\rm min}^{4q}$ from below. 

The pentaquark potential shows similar behaviour:
For diquark separations $R_1$ small as compared to $R_2$ the potential is
well approximated by the sum of the mesonic and baryonic potentials.
For values of  $R_1$ 
larger than $R_2$ it approaches $V_{\rm min}^{5q}$. However
in this case $V_{\rm min}^{5q}$ provides a
better description to the data than it did for the tetraquark case.
For our geometry we were able to confirm this behaviour 
for internal diquark distances  $R_2\stackrel{<}{\sim} 0.5$~fm.
For larger values of $R_2$ the data indicate that
the pentaquark potential also  approaches $V_{\min}^{5q}$
albeit with large statistical errors that need to be reduced before one
can make a definite statement. When $R_2 \stackrel{<}{\sim} 0.5$~fm the
pentaquark potential approaches  $V_{\rm min}^{5q}$ from above.
When $R_2\stackrel{>}{\sim} 0.5$~fm 
the pentaquark potential  
approaches $V_{\rm min}^{5q}$ 
from below 
at least 
for distances $R_1$ up to $\sim 1$~fm studied in this work.
Although the exact values for the distances $R_1$ and $R_2$ where
this behaviour is observed may depend on the geometry, the general
behaviour should not be affected.

In summary, these observations suggest that, when
the separation between diquarks is larger than 
the internal diquark distances with the latter obtaining 
values up to about  half a fermi, 
the tetraquark and pentaquark systems behave as multiquark
bound states rather than break into  mesons or baryons.
Although  this  behaviour is determined using a particular
geometry it should hold for general geometries that allow
diquark formation.

{\bf Acknowledgments:} We thank H. Neuberger for encouraging us to look 
at the static pentaquark potential. 

{\bf Note Added}: After the submission of this work  
we notice a preprint on the static 
potential on hep-lat~\cite{japan2}.


\begin{thebibliography}{99}
\bibitem{experiment}
LEPS collaboration, T. Nakano     {\it et al.}, Phys. Rev.  Lett.  {\bf 91},   012002 (2003);
DIANA collaboration, V. V. Barmin {\it et al.}, Phys. Atom. Nucl.  {\bf 66},   1715   (2003);
CLAS collaboration, S. Stepanyan  {\it et al.}, Phys. Rev.  Lett.  {\bf 91},   252001 (2003); 
SAPHIR collaboration, J. Barth    {\it et al.}, Phys. Lett.        {\bf B572}, 127    (2003).
\bibitem{DPP} D. Diakonov, V. Petrov and M. Polyakov, Z. Phys. A {\bf 359}, 305 (1997). 
\bibitem{lattice1} F. Csikor, Z. Fodor, S.D. Katz, T.G. Kovacs JHEP 0311, 
070 (2003);  T.-W. Chiu, T.-H. Hsieh hep-ph/0403020; 
S. Sasaki, Phys. Rev. Lett. {\bf 93} 152001(2004).
\bibitem{latt04} C. Alexandrou, G. Koutsou and A. Tsapalis, XXII International
Symposium on Lattice Field Theory (Lattice 2004), 
Fermilab, USA, June 21-26, 2004,  hep-lat/0409065.
\bibitem{lattice2} N. Mathur {\it et al.}, Phys. Rev. D {\bf 70}, 074508 (2004);
  N. Ishii, {\it et al.}, hep-lat/0408030.
\bibitem{nussinov} A. Casher and S. Nussinov, Phys. Lett. {\bf B578}, 124 (2004).
\bibitem{diamond}  X.-C. Song and S.-L. Zhu, hep-ph/0403093.
\bibitem{Jaffe} R. Jaffe and F. Wilczek, Phys. Rev. Lett. {\bf 91}, 232003 (2003).
\bibitem{tetraq} R. Jaffe, Phys. Rev. D {\bf 15}, 267 (1977); 
C.-K. Chow, Phys. Rev. D {\bf 51}, 6327 (1995);
B. Gelman and S. Nussinov, Phys. Lett. B {\bf 551}, 296 (2003).
\bibitem{xi} NA49 collaboration, C. Alt {\it et al.}, Phys. Rev. Lett. {\bf 92}, 042003 (2004).
\bibitem{multihit} G. Parisi, R. Petronzio and F. Rapuano, Phys. Lett. B {\bf 128}, 418 (1983).
\bibitem{onelink} Ph. de Forcrand and C. Roiesnel, Phys. Lett. B {\bf 151}, 77 (1985).
\bibitem{Bali} G.S. Bali, Ch. Schlichter and K. Schilling, Phys. Rev. D {\bf 51}, 5165 (1995).
\bibitem{APE} APE Collaboration, M. Albanese {\it et al.}, Phys. Lett. B {\bf 192}, 163 (1987).
\bibitem{variational} N. A. Campbell, A. Huntley and C. Michael, Nucl Phys.
{\bf B306}, 51 (1988); M. L\"uscher and U. Wolff, Nucl. Phys. {\bf B339}, 222 (1990);
 M. Guagnelli, R. Sommer and H. Wittig, Nucl. Phys. {\bf B535}, 389 (1998).
\bibitem{AFT} C. Alexandrou, Ph. de Forcrand and A. Tsapalis, Phys. Rev. D {\bf 65}, 054503 (2002);
Nucl. Phys. B (Proc.Suppl.) {\bf 106}, 403 (2002);  Nucl. Phys. (Proc.Suppl.)
 {\bf 109A}, 153 (2002).
\bibitem{connection} NERSC archive, G. Kilcup {\it et al.}, hep-lat/9609006.
\bibitem{CKP} J. Carlson, J. Kogut and V. R. Pandharipande, Phys. Rev. D {\bf 27}, 233 (1983).
\bibitem{Cornwall} J. M. Cornwall, Phys. Rev. D {\bf 54}, 6527   (1996). 
\bibitem{Cornwall2}J. M. Cornwall, Phys. Rev. D {\bf 69}, 065013 (2004).
\bibitem{Sommer} R.Sommer and J.Wosiek, Phys. Lett. B {\bf 149}, 497 (1984); Nucl. Phys. {\bf B267}, 531 (1986).
\bibitem{Bali2} G. S. Bali, Phys. Rep. {\bf 343}, 1 (2001).
\bibitem{AFJ} C. Alexandrou, Ph. de Forcrand and O. Jahn,  Nucl. Phys. B (Proc.Suppl.) {\bf 119}, 667 (2003).
\bibitem{japan} T. T. Takahashi, H. Matsufuru, Y. Nemoto and H. Suganuma, Phys. Rev. Lett. {\bf 86}, 18 (2001); 
T. T. Takahashi, H. Suganuma, Y. Nemoto and H. Matsufuru, Phys. Rev. D {\bf 65}, 114509 (2002).
\bibitem{map}  H.Ichie {\it et al.}, Nucl. Phys. {\bf A721}, 899 (2003); V. G. Bornyakov {\it et. al}, hep-lat/0401026.
\bibitem{japan2} F. Okiharu, H. Suganuma and T. T. Takahashi, hep-lat/0407001.
\end{thebibliography}
\end{document}